\documentclass[10pt]{article}
\usepackage{graphicx,xcolor,amssymb}
\usepackage[titletoc,toc,title]{appendix}
\usepackage{rotating}
\usepackage{pdflscape}
\usepackage{array}
\usepackage{amsmath}

\usepackage{bm}
\usepackage{lipsum}
\usepackage{listings}
\usepackage{caption}
\definecolor{mygreen}{rgb}{0,0.6,0}
\definecolor{mygray}{rgb}{0.5,0.5,0.5}
\definecolor{mymauve}{rgb}{0.58,0,0.82}
\definecolor{gris245}{RGB}{245,245,245}
\definecolor{olive}{RGB}{50,140,50}
\definecolor{brun}{RGB}{175,100,80}
\definecolor{dkgreen}{rgb}{0,0.6,0}
\definecolor{mauve}{rgb}{0.58,0,0.82}
\definecolor{black}{rgb}{0.0, 0.0, 0.0}
\definecolor{beaver}{rgb}{0.62, 0.51, 0.44}
\definecolor{columbiablue}{rgb}{0.61, 0.87, 1.0}
\definecolor{lightpink}{rgb}{1.0, 0.71, 0.76}
\definecolor{amber}{rgb}{1.0, 0.75, 0.0}
\definecolor{amber(sae/ece)}{rgb}{1.0, 0.49, 0.0}
\definecolor{aquamarine}{rgb}{0.5, 1.0, 0.83}
\definecolor{cadmiumred}{rgb}{0.89, 0.0, 0.13}
\definecolor{cambridgeblue}{rgb}{0.64, 0.76, 0.68}
\definecolor{electricblue}{rgb}{0.49, 0.98, 1.0}
\definecolor{paleplum}{rgb}{0.8, 0.6, 0.8}

\usepackage{pgffor} 

\usepackage{forloop} 

\begin{document}
\title{Exploring the energy landscape of the Thomson problem: local minima and stationary states}
\author{Paolo Amore \\
\small Facultad de Ciencias, CUICBAS, Universidad de Colima,\\
\small Bernal D\'{i}az del Castillo 340, Colima, Colima, Mexico \\
\small paolo@ucol.mx  \\
\and
Victor Figueroa \\
\small Facultad de Ciencias, Universidad de Colima,\\
\small Bernal D\'{i}az del Castillo 340, Colima, Colima, Mexico \\
\small vfigueroa6@ucol.mx  \\
\and
Enrique Diaz \\
\small Sibatel Communications \\ 303 W Lincoln Ave No.140, Anaheim, CA 92805 , U.S.A. \\
\small alcmaeon@gmail.com \\
\and
Jorge A. L\'opez \\
\small Department of Physics, University of Texas at El Paso, \\
\small El Paso, Texas 79968, U.S.A. \\
\small jorgelopez@utep.edu  \\
\and
Trevor Vincent \\
\small Sibatel Communications \\ 303 W Lincoln Ave No.140, Anaheim, CA 92805, U.S.A.  \\
\small trevor.j.vincent@gmail.com}

\maketitle
\begin{abstract}
We conducted a comprehensive numerical investigation of the energy landscape of the Thomson problem  for systems up to $N=150$. Our results show the number of distinct configurations grows exponentially with $N$,  but significantly faster than previously reported. Furthermore, we find that the average energy gap between independent configurations at a given $N$ decays exponentially with $N$, dramatically increasing the computational complexity for larger systems. Finally, we developed a novel approach that reformulates the search for stationary points in the Thomson problem (or similar systems) as an equivalent minimization problem using a specifically designed potential. Leveraging this method, we performed a detailed exploration of the solution landscape for $N\leq24$ and estimated the growth of the number of stationary states to be exponential in $N$.
\end{abstract}

\maketitle

\section{Introduction}
\label{intro}

Given $N$ charges on the sphere, interacting via the Coulomb potential, one may ask what is the configuration of lowest energy (global minimum of the electrostatic potential): this problem is known as the Thomson problem, a name that alludes to the atomic model put forward a long time ago by J.J. Thomson~\cite{Thomson1904}. The original motivations that lead Thomson to formulate his model have since disappeared but the problem itself remains open and it has attracted the attention of many researchers over the years.  Early on,  F\"oppl \cite{Foppl12} studied the stability of the arrangements discussed by Thomson up to $14$ charges and suggested empirical rules for the stability of these solutions. This problem was later revisited  by Cohn~\cite{Cohn56} and by Goldberg~\cite{Goldberg69}.  

Similar models, but for points repelling with different laws have also been introduced: for example, Tammes, a botanist, introduced what is now known as
{\sl Tammes problem}~\cite{Tammes30}, corresponding to finding the configurations of $N$ points on a sphere that maximize the minimum distance between any two points (essentially circle packing on a sphere); Fejes Toth considered the problem of maximizing the sum of mutual distances for $N$ points on a sphere~\cite{FejesToth56}. 

The first numerical explorations of the Thomson problem were carried out by Erber and Hockney in ref.~\cite{Erber91, Erber95,Erber97}  and by  Glasser and Every~\cite{Glasser92}; Rakhmanov, Zhou and Saff~\cite{Saff94} considered the cases of charges interacting with Coulomb potential, both with 
logarithmic potential and  maximizing the sum of reciprocal distances, for systems of up to $200$ points. Bergersen,  Boal and Palffy-Muhoray~\cite{Bergersen94} studied the case of logarithmic interactions for up to $N=65$ charges, establishing the interesting property that in this case the minimum configurations have vanishing dipole moment.

One striking aspect of Thomson (and similar) problems is the contrast between its apparent simplicity (after all, finding a generic solution is relatively simple) and 
the difficulty of obtaining rigorous results: proofs of optimality exist only  for small configurations~\cite{Schwartz13,Schwartz20}, whereas for large $N$ (number of points) 
it has been possible to establish the leading asymptotics of the total energy for the general case of Riesz potentials, which includes Thomson's problem as a special case~\cite{Saff94b,Saff97,Saff19}. 

In his list of 18 mathematical problems for the next century, Smale~\cite{Smale98} included the question (problem 7) of whether it is possible to find 
algorithmically configurations of points sufficiently close to the global minimum of the energy, for $N$ points on the 2-sphere interacting via a logarithmic potential.  More explicitly: if $\left\{ {\bf x}_1, \dots, {\bf x}_N \right\}$  is a configuration of $N$ points on the 2-sphere and 
$V_N(x) = -\sum_{i=2}^N \sum_{j=1}^{i-1} \ln \left| x_i -x_j\right|$ its energy ($\left|x_i-x_j\right|$ is the euclidean distance between any two points), one would like to  find
$\left\{ {\bf x}_1, \dots, {\bf x}_N \right\}$ such that 
\begin{equation}
V_N(x) - V_N \leq c \ln N \ ,
\end{equation}
where $c$ is a constant and $V_N$ is the energy of the  global minimum (it is worth pointing out that the problem is still open, see  \cite{Beltran11}). 
  
One of the aspects that emerged from the first numerical studies has been the challenge posed by finding the global minimum of 
a system: for example, Altschuler and collaborators~\cite{Altschuler94} proposed a method of global optimization, applying it to find the energies of selected  configurations for the Thomson problem, up to $N = 83$. Soon later, Erber and Hockney~\cite{Erber95} improved some of the new configurations in ~\cite{Altschuler94}, thus proving that they were not global minima. Erber and Hockney also pointed out that the number of local minima  of Thomson problem grows exponentially with $N$: "If this trend is sustained for larger values of N, identifying global minima among a large set of nearly degenerate states for complex systems of this type will pose formidable technical challenges"~\cite{Erber95}. In a different context, the study of packing structures in condensed matter, Stillinger and Weber~\cite{Stillinger84}  had previously remarked that the number of local minima is expected to grow exponentially with $N$.

For this reason considerable effort has been put in devising algorithms to find the global minimum of this class of problems~\cite{Morris96, Altschuler97, Xiang97}; 
even so, the computational  complexity of the problem increases quite steeply with $N$, and dealing with configurations with thousands of points is extremely difficult. For this reason  Perez-Garrido, Dodgson and Moore~\cite{Moore97} have restricted their search to solutions with the symmetry of the icosahedron (for $N=5792$ and $5882$). This  approach was later extended by  Perez-Garrido and Moore in ref.~\cite{Moore99} to even larger configurations $N$ (the largest value considered is $N=15282$).

More recently Wales and collaborators~\cite{Wales06, Wales09} have applied the basin-hopping method~\cite{Li87,Wales97,Wales99,Wales03} to find putative global minima for a large set of values of $N$, up to $N \leq 4352$. These configurations were obtained  with significant computational effort, using the GMIN program~\cite{gmin} (the configurations of the best minima for the Thomson problem over different values of $N$, including the ones of \cite{Wales06,Wales09}, are available at the Cambridge database~\cite{Walesrepo1,Walesrepo2}).  Lai and collaborators~\cite{Lai24} have recently studied the Coulomb and logarithmic potentials for configurations up to $N=500$ charges on the sphere, using a population-based heuristic approach
and obtaining encouraging results.

A related but broader problem than just finding the global minimum of Thomson problem is finding {\sl all} the local minima for a given $N$.  As we mentioned earlier,
Erber and Hockney~\cite{Erber97} carried out an exploration for $N \leq 112$, confirming their earlier finding that the number of minima grows exponentially.

More recently, Calef et al.~\cite{Calef15} considered a generalization of the Thomson problem, in which the charges interact with a Riesz potential 
\begin{equation}
V(r) = \left\{   \begin{array}{ccc}
r^{-s}  & , &  s>0\\
- \log r &,& s=0
\end{array}\right. \ ,
\label{eq_potential}
\end{equation}
where $r$ is the distance between two charges and $s$ is a parameter ($s=1$ corresponds to the Coulomb potential). 
The original goal of \cite{Calef15} was to identify the majority  of local minima for  $N\leq 180$  and  for $s=0,1,2,3$ but after intensive
numerical explorations   they noticed that the number of new configurations found was not decaying sufficiently fast, suggesting  that the actual number of local minima may be much larger 
than what observed. To avoid this problem they described a procedure to estimate the true number of local minima for a given $N$. 
Fig.~6 of \cite{Calef15} shows a comparison between the observed and  estimated configurations for the case of the Coulomb potential: for the largest case discussed by those authors, $N=180$, 
they estimated $\approx 30000$ configurations, while finding less than $\approx 3000$ (the exact number of configurations found in \cite{Calef15} is not reported in the paper and therefore approximated values have to be extracted from the plots).

In a later paper, Mehta and collaborators~\cite{Mehta16} studied the Thomson problem ($s=1$) for selected values of $N$ ($N=132, 135,138,141,144,147,150$): their numerical exploration led them to conclude that the energy landscape of the Thomson problem,  at least for the values of $N$ considered in  their work is single funneled, a characteristic that  is generally associated with small world characteristics~\cite{Doye02}. Under these circumstances the search for the global  minimum is easier. This aspect is relevant in connection to Smale's 7th problem~\cite{Smale98}.

In particular they found that the number of local minima for the cases considered exceeds greatly the ones (both observed and  estimated) found in \cite{Calef15}.   
Additionally they also  searched the transition states, defined as configurations with vanishing gradient and one negative eigenvalue of the Hessian matrix (see Fig.~1 of \cite{Mehta16} for a comparison with the results of \cite{Calef15}). 
The results of Mehta et al.  justify the expectation that a similar behavior even for values of $N$ not considered in \cite{Mehta16}, with  
a number of local minima much larger that what expected from the fits of refs.~\cite{Erber91,Erber95,Erber97,Calef15}.

In this paper we want to address two different aspects in connection to the Thomson problem: the first one is to perform a more careful exploration of the energy landscape 
(in other words, trying to identify the majority of local minima for a given $N$) over a wider range of values  (we have performed a systematic search for  $N \leq 150$ and partial searches for  $N=180$)  and use the results to obtain a more precise estimate of the exponential growth with $N$ of the local minima;  the second one is to perform an exploration of the solution landscape (i.e. the set of all stationary points for a given $N$)  for small values of $N$ ($N \leq 24$).
Wales and Doye~\cite{Wales03b} have found that the number of stationary states for a high--dimensional system that can be subdivided into  a number of independent subsystems should follow essentially a normal distribution in the index (number of negative eigenvalues of the Hessian). 
Curiously, we have found only two references where selected stationary states for the Thomson problem have been calculated: 
ref.~\cite{Mehta15} and ref.~\cite{Yin22}, respectively using the Newton homotopy method and a constrained high-index saddle dynamics (CHiSD) method to find 
stationary points corresponding to a range of indices.  In particular  Yin and collaborators in Fig.2 of  ref.~\cite{Yin22} report a graphical 
representation of the solution landscape for $N=5,7,9$. Our approach to this problem consists of transforming the search for stationary states into
a minimization problem, introducing a suitable function (not just the total energy of the system).

Both tasks considered in the present paper have represented a strong computational challenge that has required intensive numerical calculations.

The paper is organized as follows: in section \ref{sec:num} we describe the different computational strategies that we have used to attack these problems; in section \ref{sec:results} we report the numerical results (the subsections \ref{sec:loc_min} and \ref{sec:stat_states}  contain the cases of local minima and stationary states, respectively). Finally in section \ref{sec:concl} we draw our conclusions.

\section{Computational strategies}
\label{sec:num}

We consider $N$ points on a sphere interacting with a potential in eq.~(\ref{eq_potential}), where $s=1$ corresponds to the Thomson problem studied here.
The total energy of the system 
\begin{equation}
E = \sum_{i=2}^N \sum_{j=1}^{i-1} V(r_{ij})  \ ,
\label{eq_energy}
\end{equation}
where $r_{ij}$ is the euclidean distance between a pair of charges, is  a function of $2N-3$ angles~\footnote{Due to the symmetry of the sphere one can eliminate three degrees of freedom from the problem by assigning a charge to the north pole and a second charge to the $xz$ plane.}. We are interested in calculating the local minima and stationary configurations of eq.~(\ref{eq_energy}) for a given number of charges repelling via the Coulomb potential.

\subsection{Local minima}

The approaches that we have followed to find the local minima of eq.(\ref{eq_energy}) are:
\begin{itemize}
\item {\sl straightforward approach}

Given $N$ random points on a sphere, a direct minimization of eq.~(\ref{eq_energy}) is performed, eventually leading to a local minimum of the total energy; 
this procedure is repeated many times and the configurations obtained in this way are then filtered, eliminating duplicate configurations. 
In general we have used this approach only to find the first few instances of local minima for a given $N$, since the computing time required to 
reach a minimum is typically larger than in other approaches (since the initial configurations are completely random more iterations are needed
to converge to a solution). Because the minima are not equally probable, generating the majority of local minima for a given $N$ may be very difficult 
using this approach, unless $N$ is sufficiently small.

\item {\sl perturbation approach}

Given a configuration corresponding to a local minimum of eq.~(\ref{eq_energy}), the positions of the points are perturbed by applying a random noise, followed by a minimization of the energy functional; a single original configuration is perturbed several times and the resulting configurations generated in this way are then filtered to eliminate duplicates.  

The size of the perturbation is initially set to some small value which is increased when no further independent configurations are found (usually 2-3 times);  
notice that setting the size of the perturbation either too small or too large would result most likely either in falling back on the original configuration or in missing configurations which are close to the original. Additionally  the computing time is sensitive to this parameter, since larger perturbations require more iterations to reach convergence. Typically for the first stage we have performed $50$ perturbations for each original configuration, while for the second stage (where the size of the perturbation is increased), we have used a much larger number (the further perturbations are applied only to newly found configurations).
This cycle is repeated a large number of times ($\approx 100$). For the case of the largest $N$ considered in this paper, a complete scan using this approach 
can produce several millions of configurations (before the filtering process).

\item {\sl controlled perturbation approaches}

The amount of chaos introduced by a random perturbation affects the computing times in two ways: first because the gradient can be sizable for most of the points,  
given the disorder of the initial configuration,  second because the different ansatzes obtained by perturbing the same configurations may not be sufficiently different 
to produce as many independent configuration as possible.

We have implemented three different approaches that allow to avoid this problem, using configurations that are local minima of the problem:

\begin{itemize}
\item {\sl point migration}

Consider a configuration of $N$ points on the sphere that is a local minimum of the energy, eq.~(\ref{eq_energy}), and calculate the corresponding  Voronoi
diagram: an ansatz can be produced by moving $N_0$ of the $N$ points along the spherical segments that go though the original position  and
one of the vertices of the Voronoi cell, without exiting the cell (we call $\alpha$ the parameter controlling the displacement). Using this procedure each configuration can generate a large number of ansatzes, where the points are still well separated and the gradient is small over a large set of points (because the perturbation is localized, points that are far from the location where the perturbations have been applied will be almost in equilibrium).  By varying the parameters that control the amount of perturbation introduced in the configuration, $N_0$ and $\alpha$, one can thus produce independent ansatzes quite efficiently: these ansatzes, are then minimized and the configurations produced in this way are then filtered to retain the independent configurations.

\item {\sl downgrade algorithm}

Given a configuration with $N+1$ points corresponding to a local minimum of the total energy, an ansatz for a configuration with $N$ points is obtained by eliminating one of the points and then by minimizing the total energy; in this case, the positions of the points that are retained are unchanged and again one expects that points that are farther from the region where the removal occurred  to be almost in equilibrium. Due to the absence of random perturbations, the outcome is deterministic and the algorithm needs to be applied just once to a set of configurations. In this framework each configuration with $N+1$ points  produces $N$ ansatzes. 

\item {\sl upgrade algorithm}

Given a configuration with $N-1$ points corresponding to a local minimum of the total energy, a new ansatz for $N$ points can be obtained by introducing  an extra point, located at one of the vertices of the Voronoi diagram. This operation is done for {\sl all} the points and all the vertices of the Voronoi diagram, resulting in $2N-4$  ansatzes~\footnote{We are assuming that the Voronoi diagram contain only vertices with three legs; for the Thomson problem, this is overwhelmingly the case.}. This algorithm bears some similarity with the one described in section 2.3 of ref.~\cite{Cheviakov18}, with the  difference that the new point is obtained from the Delaunay diagram of the original configuration and  corresponds to the center of mass of a particular cell (which is not necessarily a vertex of the Voronoi diagram).

Because the upgrade and downgrade algorithms are deterministic, at each stage only the configurations that have not previously been   acted upon are processed; in this way one is able to  avoid the {\sl avalanche } effect seen when applying random perturbations and makes our approach particularly efficient for larger $N$.
Clearly the simplest version of this algorithm corresponds to $N_i = N$ and $N_f=N+1$.

\end{itemize}

\end{itemize}

Our normal course of action has consisted of first applying the straightforward approach with a limited number of trials, followed by the remaining 
approaches, in different order and repeatedly. In this way we have been able to improve all the results for the local minima of the Thomson problem 
reported by Mehta and collaborators in  \cite{Mehta16} (a discussion of the numerical result is presented in the next section). We have found that the 
upgrade/downgrade algorithms are particularly efficient in producing new configurations, particularly for the largest $N$ considered in the present paper.

Another aspect that is crucial in our exploration is the minimization method adopted, because of the simultaneous need for both speed and precision. We have found that 
the best performance was obtained using the Newton-conjugate gradient method, using the explicit representation for the Hessian matrix. All programs were implemented
in python~\cite{python}, using the scipy library~\cite{scipy}  and whenever possible using numba acceleration~\cite{numba}  and running in parallel~\footnote{
As a technical note we have found that the different minimization methods available in scipy that we have tested have different sensitivity to the hardware. 
In particular  the Newton-conjugate method is quite efficient on the processors AMD Ryzen 9, 7950X × 32 that we had at our disposal.}. 

Although we have mostly run on CPU a partial adaptation of the program has
been done for GPU and used to calculate a small portion of the conﬁgurations
for the largest N considered in this paper, N = 180. The GPU code was developed using PyTorch and pytorch-minimize, employing the Newton-Conjugate solver consistent with the CPU implementation. To further accelerate computations, we leveraged Multi-Instance GPU (MIG) support on H100 cards to parallelize the workload.

Most of the configurations produced in  the minimization stage, which at least for the largest $N$ considered in this work may amount to few millions, 
are not independent and therefore they need to be filtered,  retaining only those that are effectively different.  This filtering is performed on the basis of energy comparison, rejecting configurations with energies differing  by an amount less than the tolerance (usually we have set this tolerance at $10^{-8}$). Configurations that don't pass this filter may in principle still be different and an additional filtering process may be needed: we 
postpone the discussion of this procedure to subsection \ref{sub_comp}.

We have observed that the configurations calculated with the methods discussed above have at least $12$ significant decimals in the energy~\footnote{In the calculation of the energy we have applyied the Kahan summation algorithm to increase the accuracy of the results.} and a norm of the gradient which is small  (the typical order of magnitude is $10^{-9}$) . However, finding the most accurate numerical results for the minimal  is essential, particularly in order to assess whether the solution found is effectively a local minima or a stationary point with higher index. For this reason these solutions undergo a refinement stage in which they are passed through the Newton's method reaching gradients with norms typically of  size $~\approx 10^{-13}$ (much smaller values of the gradient could be obtained working in arbitrary precision, but for the sizes of $N$ considered here it would be impractical and time consuming).

After completing the refinement stage, a certification stage follows: in this procedure   the configurations that have made it to this point  are checked one by one, verifying that all the eigenvalues of the hessian are positive or null  (within some numerical tolerance), i.e. that the configuration corresponds to a local minimum of the energy.  In general we have found that occasional stationary states are extremely rare. 

\subsection{Stationary configurations}
\label{sub_st}
Let us now discuss the second task of this  paper, namely finding the stationary configurations for a given $N$.  These configurations 
have a vanishing gradient but can be either local minima, saddle points or maxima of the total energy functional.
Although our numerical results will be limited to the Coulomb potential ($s=1$) we keep our discussion general by considering an arbitrary $s$.

The total electrostatic energy is 
\begin{equation}
\begin{split}
E(s) = \left\{
\begin{array}{ccc}
\frac{|s|}{s}  \sum_{i=2}^N \sum_{j=1}^{N-1} \frac{1}{r_{ij}^s} & , & s \neq 0 \\
\sum_{i=2}^N \sum_{j=1}^{N-1} \log \frac{1}{r_{ij}} & , & s = 0 \\
\end{array}
\right. 
\end{split}
\label{eq_V}
\end{equation}
where \begin{equation}
r_{ij} = 2 \left[ 1- \cos \left(\theta_i\right)  \cos \left(\theta _j\right)
- \ \cos \left(\phi_i-\phi _j\right) \sin \left(\theta_i\right) \sin \left(\theta _j\right) \right]
\end{equation}
is the euclidean distance between two charges. The total energy  is a function of the $2N$ angles ( as mentioned earlier, the degrees of freedom
can be reduced to $2N-3$  when the symmetries of the problem are taken into account).

Because the configurations that we are looking for are not just local minima or maxima of eq.~(\ref{eq_V}), we need to define an alternative 
functional to work with. Let $\vec{f}_{ij}(s) \equiv  \frac{\vec{r}_{ij}}{r_{ij}^{s+2}}$ be the repulsive force between two equal charges; 
the total force on the  $i^{th}$ charge is
\begin{equation}
\vec{F}_i(s) =  \sum_{j \neq i}  \vec{f}_{ij}(s) =  \sum_{j \neq i}  \frac{\vec{r}_{ij}}{r_{ij}^{s+2}}   \ .
\end{equation}

The component of this force  on a plane tangent to the sphere at $\vec{r}_i$ is obtained as
\begin{equation}
\vec{\mathcal{F}}_i(s) \equiv \vec{F}_i(s) - \left( \vec{F}_i(s) \cdot \hat{r}_i \right) \ \hat{r}_i \ .
\end{equation}

Using the electrostatic energy of the $i^{th}$ charge, $V_i = \frac{1}{2} \sum_{j\neq i} \frac{1}{r_{ij}^s}$, one can obtain
a more compact expression for $\vec{\mathcal{F}}_i(s)$, using the identity $\vec{F}_i(s) \cdot \hat{r}_i  = V_i$:
\begin{equation}
\vec{\mathcal{F}}_i(s) \equiv \vec{F}_i(s) - V_i \ \hat{r}_i \ .
\end{equation}

At equilibrium one must have
\begin{equation}
|\vec{\mathcal{F}}_i(s)| =0  \hspace{1cm} , \hspace{1cm}  i=1, \dots, N
\end{equation}
and the function
\begin{equation}
\begin{split}
\mathcal{V}(s) &\equiv \sum_{i=1}^N |\vec{\mathcal{F}}_i(s)|^2 
= \sum_{i=1}^N \sum_{j,k \neq i}^N \frac{1}{(r_{ij} r_{ik})^{s+2}} 
\left[  \left(  r_{ij}^2 - \frac{r_{jk}^2}{2}\right) - \frac{r_{ij}^2 r_{ik}^2}{4}\right]
\label{eq_functional1}
\end{split}
\end{equation}
is the generalization of eq.~(\ref{eq_V}) and the stationary configurations of eq.(\ref{eq_V}) are now {\sl degenerate} (global) minima of $\mathcal{V}(s)$.

To perform an  exploration of the minima of (\ref{eq_functional1}) we have implemented the Newton algorithm both in python and
in mpmath~\cite{mpmath}. This is a three stage process: first a minimization of (\ref{eq_functional1}) is performed in python, working in double precision (typically using the truncated Newton method in scipy);
in a second moment,  the  python version of Newton' method is applied to the configuration obtained in the first stage until sufficient precision is achieved and finally 
the mpmath version of Newton's method is applied , typically obtaining about $30$ significant digits in the energy~\footnote{In this way we are able to limit the number of costly iterations performed in multiple precision.}. 

The high accuracy of our approach is essential at the moment of filtering out duplicate configurations and also allows one to calculate very precisely the eigenvalues of the Hessian matrix, which are needed to perform a classification of the configuration (the relevant parameter here is the index,  i.e. the number of negative eigenvalues of the hessian).

\subsection{Comparison}
\label{sub_comp}

Configurations that don't pass the energy test may, at least in principle, still be different. We will discuss here two further tests
that can help to ascertain whether two configurations are different.

The first test is a generalization of the energy test discussed earlier: given two configurations with the same number of points, for each configuration we calculate the electrostatic energy of each particle, due to the interaction with all remaining particles belonging to the same configuration.  We then proceed to progressively pair each component of the energy vector for the first configuration with a component of the energy vector in the second configuration: this is done starting from the pair of components which are most similar and iterating until all components are paired. 
Finally one calculates the sum of the absolute values of the differences between components: for configurations that are very similar the magnitude of this quantity will be small.

The second test that we have implemented  essentially corresponds to algorithm 1 of ref.~\cite{Calef15}, with some difference. 
Our algorithm works in the following way:
\begin{itemize}
	\item[a)] consider two different configurations with $N$ points on the sphere and with sufficiently similar energies, $E_1$ and $E_2$; let ${\bf r}_i^{(1,2)}$ with $i=1,\dots, N$, be the points of these configurations;
	\item[b)] for each configuration construct the tables of distances with elements $d_{ij}^{(1)} = |{\bf r}_i^{(1)} -{\bf r}_j^{(1)}|$ and $d_{ij}^{(2)} =|{\bf r}_i^{(2)} -{\bf r}_j^{(2)}|$ and form the vectors $V_i^{(1,2)}  = \left[ d_{i1}^{(1,2)}, \dots, d_{iN}^{(1,2)}\right]_{\rm ordered}$, with elements written in ascending order (clearly the first element of the vector is  $d_{ii}^{(1,2)}=0$);
	\item[c)]  Now consider the norms $|V_i^{(1)} - V_j^{(2)}|$ for $i,j=1,\dots, N$ and order them starting from the pair $(i,j)$ corresponding to the smallest norm; in  this way we have a sequence of pairs $(i_1, j_1)$,  $(i_2, j_2)$, \dots and perform a rotation that brings the points $i_1$ and $i_2$ of the first configuration to the north pole and on the $xz$ axis, respectively; do the same for the second configuration;
	\item[d)]  Now that the two configurations have been rotated, we can superpose them and proceed to pair each point in the first configuration 
	with a point in the second configuration at the smallest distance, starting from the pairs with closest overlap, progressively exhausting all pairs (once two points have been paired they are no longer available for comparison); there are $N$ pairs, each with a distance $D_p$, $p = 1,\dots,N$ and one may define the quantity  $\Delta =  \sum_{p=1}^N D_p$; 
	\item[e)]  the procedures at point $c)$  have to be repeated, considering the second configuration modified by performing a mirror reflection about the $xz$ plane; in this case a different value of $\Delta$ will be obtained;
	\item[f)]  pick the smallest value of $\Delta$  between the two obtained at steps $d)$ and $e)$; the smallness of $\Delta$ is an indication of the similarity of the configurations: if its value is below the tolerance conclude that the configurations are possibly identical;
\end{itemize}

Although we have discussed these procedures in connection to the filtering process, the comparison of configurations is useful not just to discard redundant configurations, but also to assess the degree of similarity between configurations that are different. In the next section we will discuss applications of these ideas.

\section{Numerical results}
\label{sec:results}

In this section we present the numerical results obtained using the computational approaches described in the previous section. 
These results are available for download at Zenodo.

\subsection{Local minima}
\label{sec:loc_min}

The first attempt of calculating the local minima for the Thomson problem over a range of values of $N$ was performed Erber and Hockney~\cite{Erber91,Erber95,Erber97}.  In particular the numerical results reported in their paper ref.~\cite{Erber97}  were obtained running 
on the ACPMAPS supercomputer at Fermilab (a parallel processing machine with $600$ double precision cores) and reached configurations of up to $112$ points. 

More recently, Calef and collaborators~\cite{Calef15} have made an effort to extend the exploration started by Erber and Hockney both to larger $N$ ($N\leq 180$) and to  different potentials ($s=0,1,2,3$). Their ambitious plan however fell short of delivering satisfying results, as they soon realized that the number of configurations effectively observed in their numerical experiments was most likely only a tiny fraction of the  true set of local minima.   

In fact, Mehta and collaborators~\cite{Mehta16} have found that the number of local minima, $N_{\rm conf}(N)$,  for the Thomson problem at select values of $N$ is much larger that what found in \cite{Calef15}; as a  result of this the fits of $N_{\rm conf}$ reported in \cite{Calef15}  are not very accurate.
 
Based on these considerations we have focused our efforts on the Thomson problem ($s=1$) and for all configurations with $N \leq 150$, with the additional special case   $180$, which is the largest considered in \cite{Calef15} ($100 \leq N \leq 180$).  We have performed a large number of numerical trials, using the various approaches described in the previous section and we have managed to increase the values of $N_{\rm conf}(N)$ for several values of $N$, including those studied in ref.~\cite{Mehta16}. For the larger configurations considered here this has required to produce several million  trials for each $N$. 

\begin{figure}
\begin{center}
\includegraphics[width=10cm]{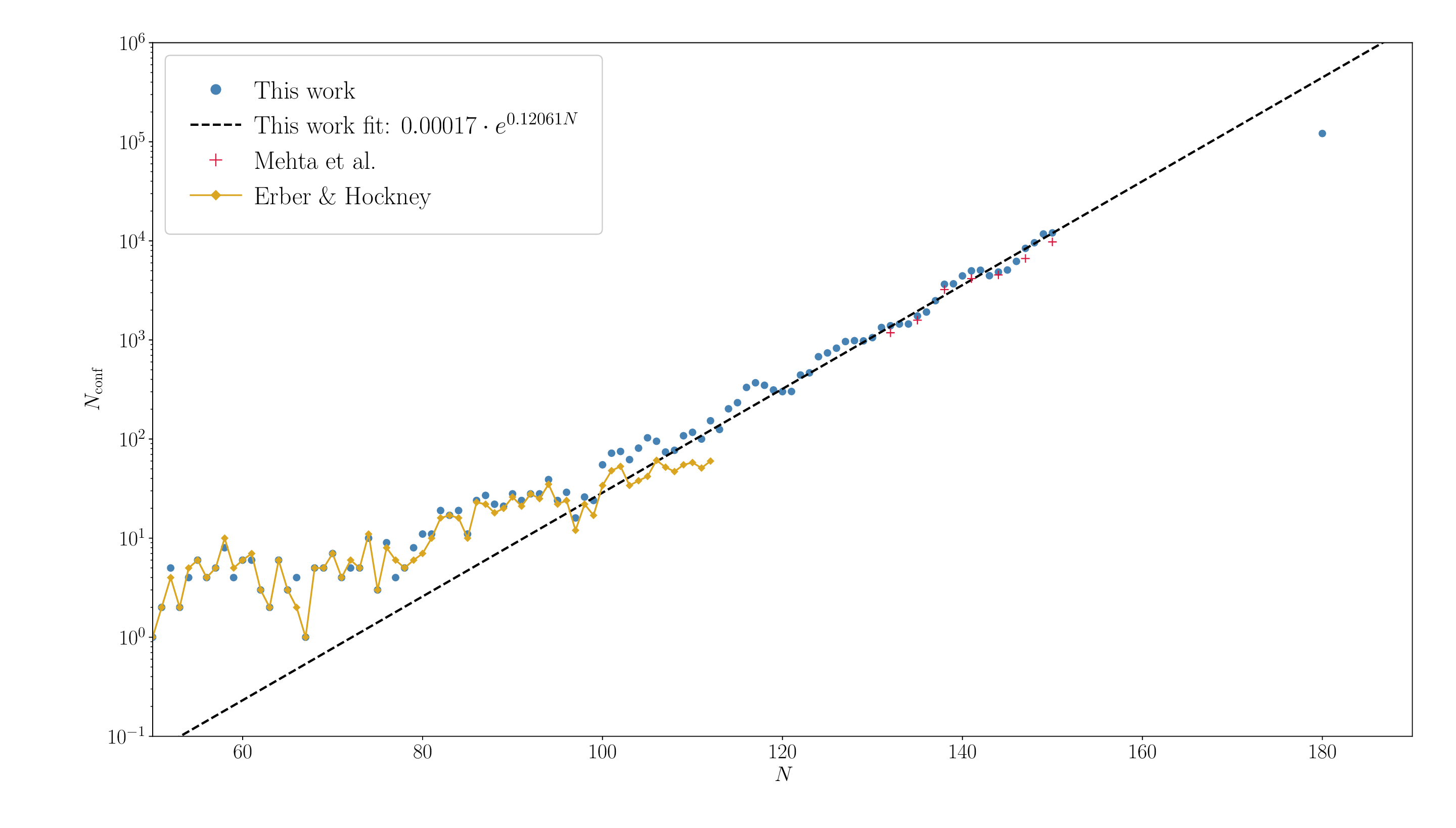}    
\caption{Number of local minima for the Thomson problem.}
\label{Fig_N_local}
\end{center}
\end{figure}

The fruit of our hard work is displayed in figure ~\ref{Fig_N_local}, where we plot the $N_{\rm conf}$ for $10 \leq N \leq 150$ 
and $N = 180$ (blue circles) and compare it with the similar results of  refs.~\cite{Erber97} (yellow diamonds) and ~\cite{Mehta16} (red crosses)~\footnote{The plot does not include a comparison with  ref.~\cite{Calef15},  because the actual number of observed local minima at selected values of $N$ was not reported there. }. 
The solid line in the plot is the exponential fit obtained using the results for $100 \leq N \leq 150$
\begin{equation}
N_{\rm conf}(N)  = 0.00017 \times e^{0.12061\ N} \ ,
\label{fit_nconf}
\end{equation}
which suggests $N(180) \approx  4.55 \times 10^5$ (more than $100$ ($10$) times the result {\sl observed} ({\sl estimated}) in Fig.6 of \cite{Calef15}) .

As mentioned in ref~\cite{Mehta16}, for $N > 400$ the global minima found in \cite{Wales06} start to display alternative defect motifs and
possibly the potential energy landscape may display multiple funnels in this regime. The fit (\ref{fit_nconf}) in this case provides
$N_{\rm conf}(400) \approx 1.5 \times 10^{17}$: it is clear that for systems of this size or larger it is {\sl impossible} to carry out a
detailed exploration as done in \cite{Mehta16} or in the present paper.

\begin{figure}
\begin{center}
\includegraphics[width=10cm]{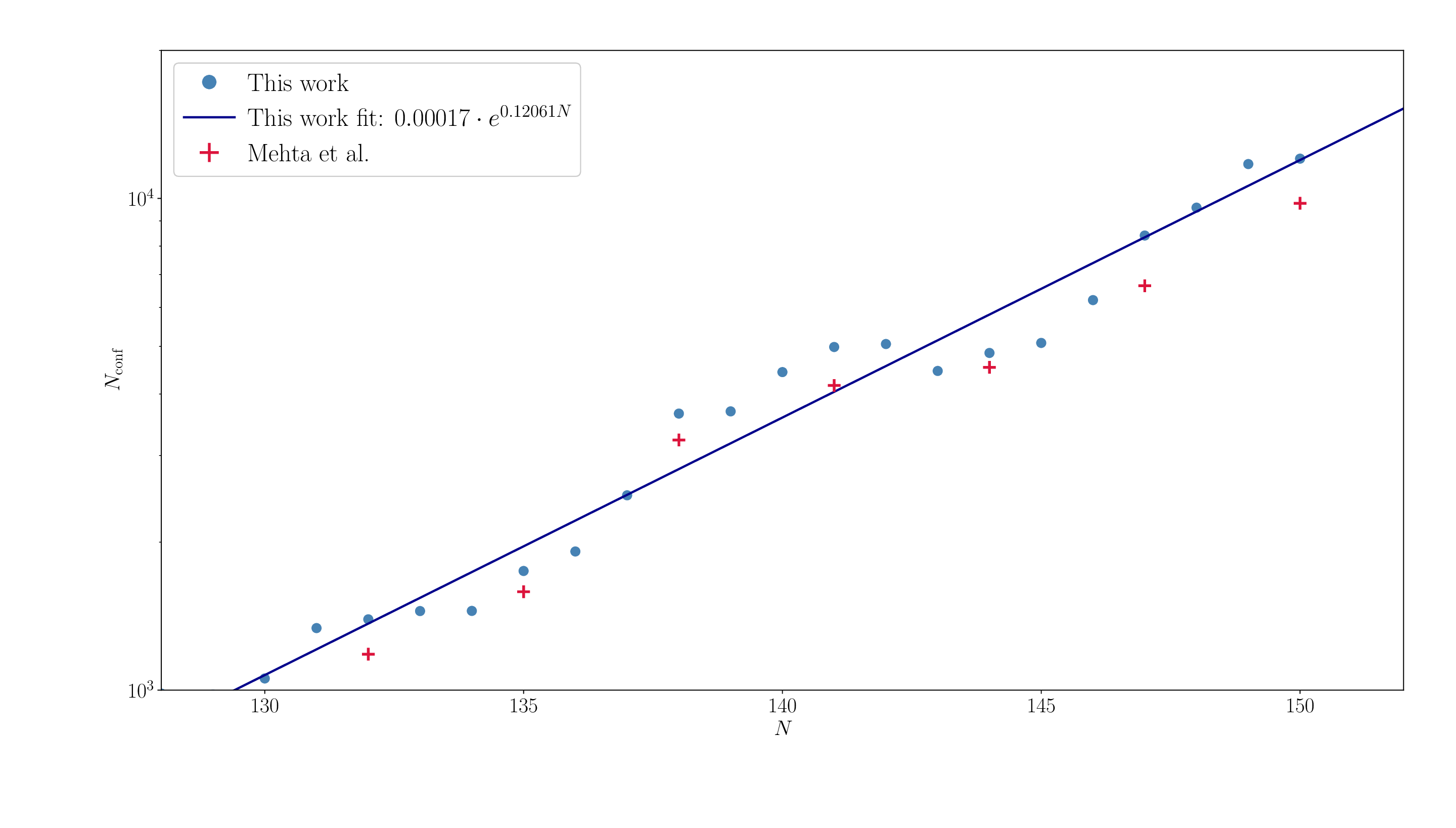}    
\caption{Number of local minima for the Thomson problem (detail).}
\label{Fig_N_local_detail}
\end{center}
\end{figure}

In Fig.~\ref{Fig_N_local_detail} we show a detail of Fig.~\ref{Fig_N_local}, over a region that contains the values of $N$ studied in  \cite{Mehta16}.
In particular,  the number of local minima found in our explorations exceeds the results of \cite{Mehta16} by sizable amounts: for $N=147$,
for example,  we have found $8356$ configurations, compared to the $6644$ of \cite{Mehta16}). 
Even if reaching the global minimum is relatively easy for systems of this size, it is evident that performing an exhaustive exploration of the energy landscape constitutes a  tremendous numerical challenge.

There is an other computational aspect that  should be discussed:  the primary  tool to  decide whether two configurations are different is the energy comparison. As a matter of fact, if the energy of two configurations is different, the configurations are necessarily different.  
As we have described in the previous section,  the comparison is performed using a tolerance, which should be small, but sufficiently larger than round--off errors that inevitably appear in a numerical calculation. If the energy difference between the two configurations is larger than the tolerance, one can safely conclude that the configurations are different, however, if it is smaller, some caution should be used.  In such case it is  not possible to decide whether two energies are different, if their difference is too small 
because of the finite numerical accuracy.

\begin{figure}
\begin{center}
\includegraphics[width=10cm]{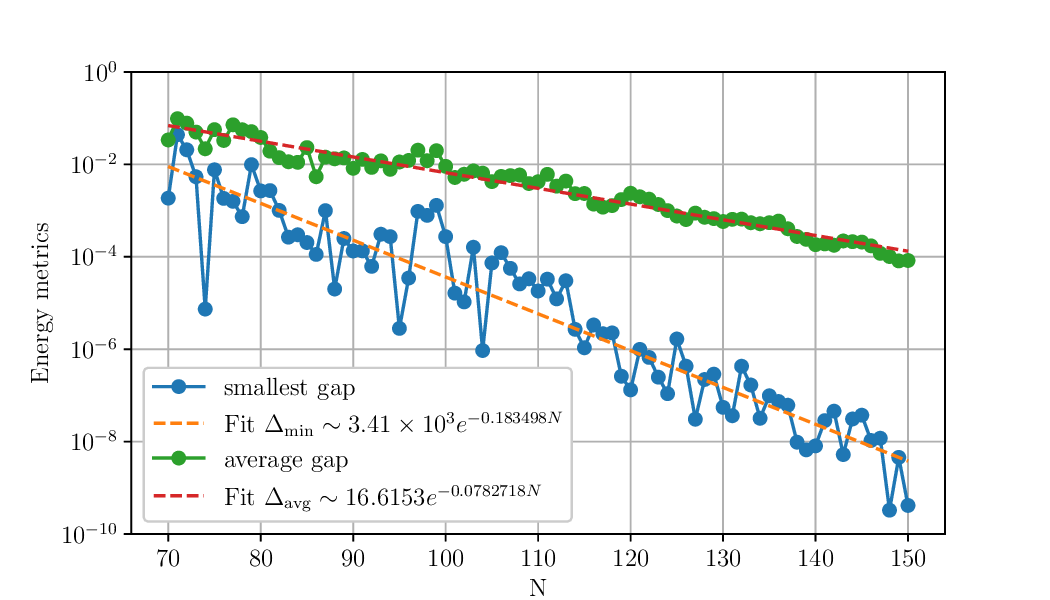}    
\caption{Smallest energy gap for configurations with $80\leq N \leq 150$.}
\label{Fig_smallest_gap}
\end{center}
\end{figure}

As first sight, it may appear that we are worrying about events that possibly are very rare and therefore should affect the exploration only marginally, if at all. To prove or disprove such expectation we have considered the configurations found for a given $N$ and sorted them in energy (from the smallest to the largest), calculating the energy gaps between consecutive configurations and selecting the smallest gap at each $N$. The results are plotted in figure \ref{Fig_smallest_gap}, where the small circles are the numerical results, whereas the dashed line is an exponential fit.

Similarly we have calculated the average gap (green dots in the figure) and found that it also decay exponentially with $N$, although more slowly: 
the fits that we have obtained are 
\begin{equation}
\begin{split}
\Delta_{\rm min} &= 3410 \times e^{-0.183498 \ N} \\
\Delta_{\rm avg} &= 16.615  \times e^{-0.0782718 \ N} \\
\end{split} \ .
\end{equation}

For $N = 400$, for instance,  these formulas provide the estimates
\begin{equation}
\begin{split}
\Delta_{\rm min} &\approx  4.5 \times 10^{-29} \\
\Delta_{\rm avg} &\approx 4.2 \times 10^{-13} \\ 
\end{split} \ ,
\end{equation}
which suggest that the gaps of a large portion of the configurations might be much smaller than the precision accessible while working in double precision~\footnote{This is an additional challenge to the full exploration of the energy landscape.}.

\begin{figure}
\begin{center}
\includegraphics[width=10cm]{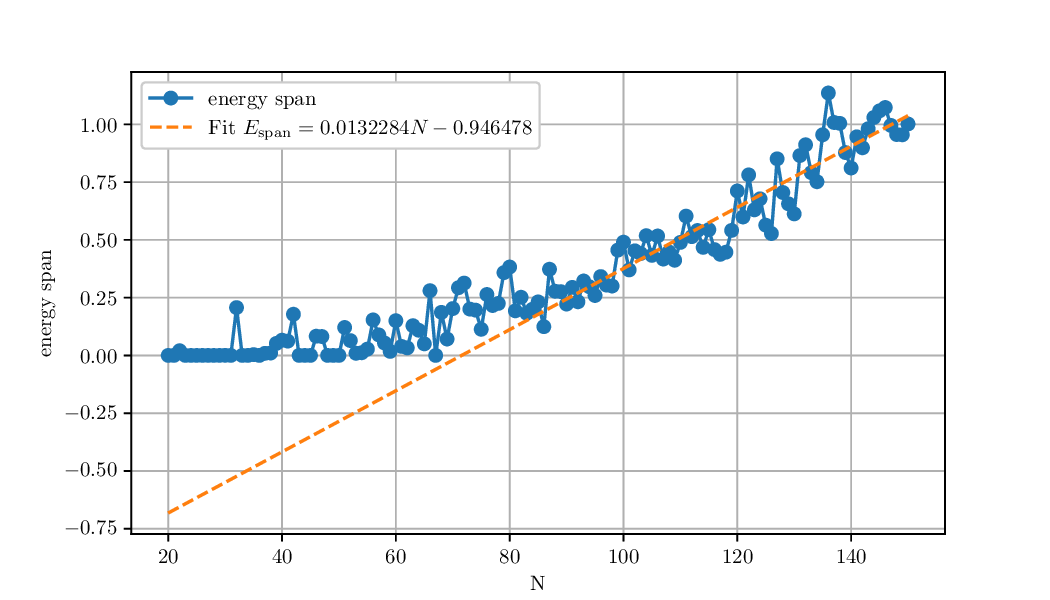}    
\caption{Energy span defined as $E_{\rm max}(N) -E_{\rm min}(N)$ for configurations with $80\leq N \leq 150$.}
\label{Fig_energy_span}
\end{center}
\end{figure}

In Fig.~\ref{Fig_energy_span} we plot the total energy span for a given $N$,  $E_{\rm max}(N) - E_{\rm min}(N)$, as  function of $N$ and find that this quantity 
is growing essentially {\sl linearly} with $N$; this behavior helps to understand why the average gap between consecutive configurations is decaying exponentially:
in fact, the number of independent configurations is increasing exponentially with $N$, but the range of energy where they accommodate is growing approximately linearly.

\begin{figure}
\begin{center}
\includegraphics[width=10cm]{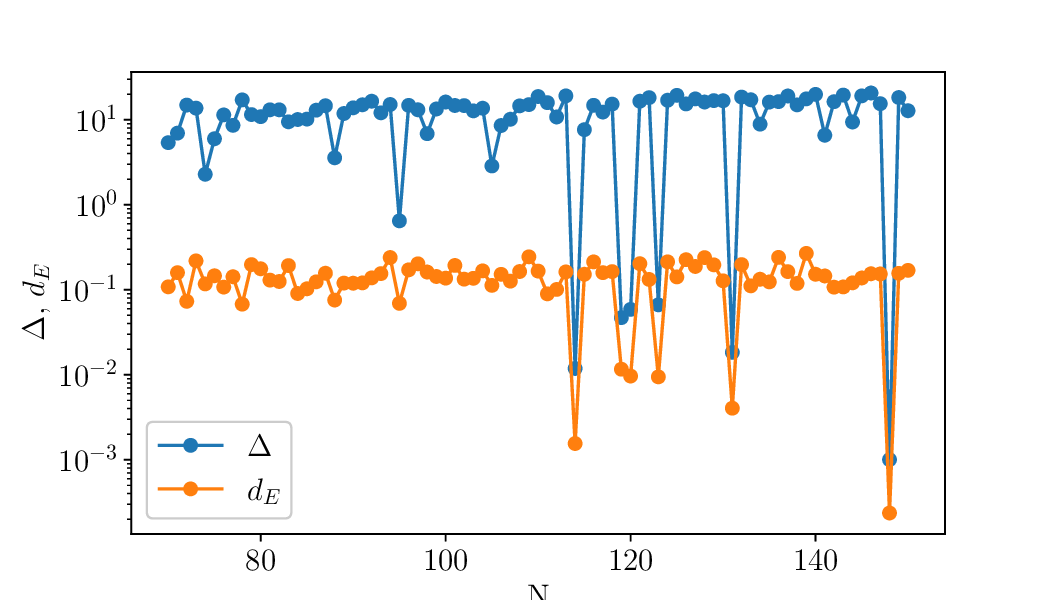}    
\caption{Smallest energy gap for configurations with $80\leq N \leq 150$.}
\label{Fig_gap_comparison}
\end{center}
\end{figure}

Because the energy gaps between configurations can be extremely small, depending on $N$, one cannot rely uniquely on the energy to resolve different configurations.
In Fig.~\ref{Fig_gap_comparison} we apply the comparison methods that we have described in \ref{sub_comp}  to the pairs of  configurations with smallest gaps
for  $70 \leq N  \leq 150$. Observe that the two comparison are in agreement.
The main information that we extract from this plot contradicts the naive expectation that configurations that are almost degenerate in energy are  very similar.

\begin{figure}
\begin{center}
\includegraphics[width=6cm]{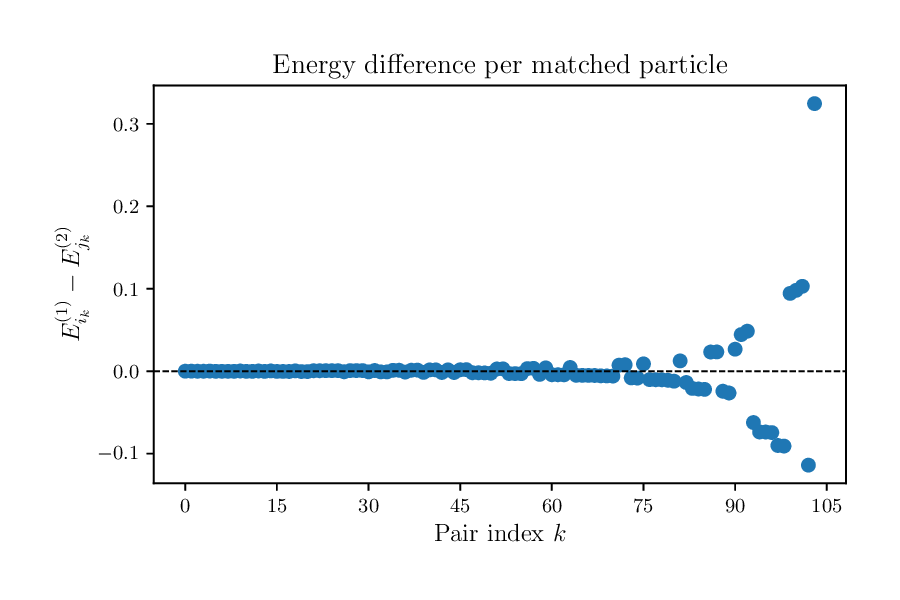}    
\includegraphics[width=6cm]{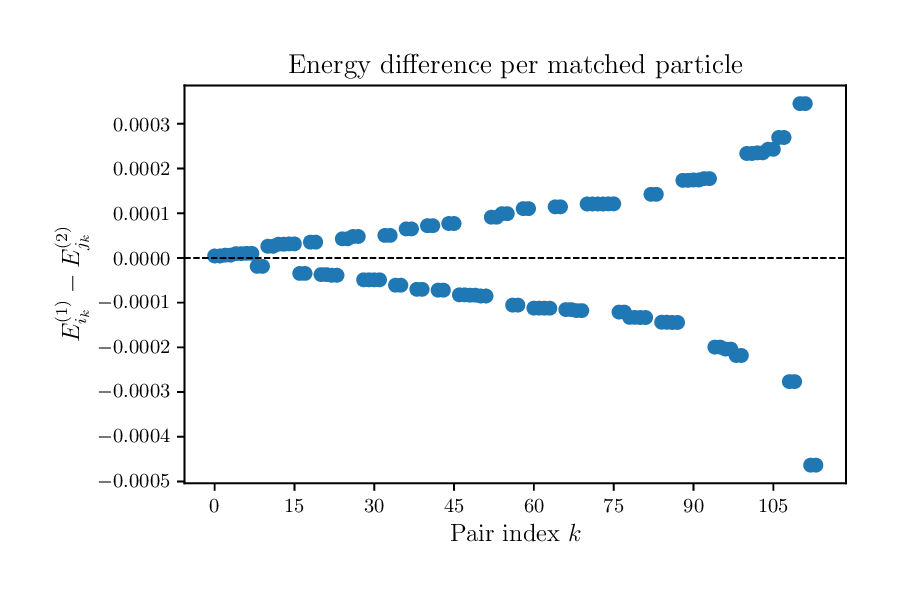}    
\caption{Energy comparison for almost degenerate configurations with $N=104$ (left plot) and $N=114$ (right plot).}
\label{Fig_energy_difference}
\end{center}
\end{figure}

In the left plot of  Fig.~\ref{Fig_energy_difference} we plot the differences between the paired energy  components of the two configurations with smallest gap  for $N=104$, with energies  $4823.130667492144$ and $4823.130668427168$. Their  gap is just $\delta E \approx 9.35 \times 10^{-7}$. This value, however, turn out to be this small not because the individual terms are themselves that small, but because of the large cancellation in the sum. The energy distance, $d_E$,  is defined as 
\begin{equation}
d_E = \sum_{i=1}^N \left|  E_{p_i^{(1)}}-E_{p_i^{(2)}}\right|  \ ,
\end{equation}
where $p_i^{(1,2)}$ are the paired elements of the two configurations at $i^{th}$ position.  For the case under consideration, $d_E \approx 1.73 \gg \delta E$ and 
we can safely conclude that the two configurations are different. Notably, the Voronoi diagrams of the two configurations are also
different, with the lowest one having $12$ pentagonal defects and the highest one with $13$ pentagons and one $heptagon$. 

In the right plot of the same figure we plot the similar quantities for the case of $N=114$, corresponding to energies $5826.5962206956465$
and $5826.596223384571$; in this case we observe that the energy distance $d_E$ is much smaller, signaling that the two configurations, while different, are 
much more similar (indeed in this case the same number of defect is found on the two): this situation is analogous to the one observe in  ref.~\cite{Calef15}, where it was found out that two configurations for $N=102$ and $s=2$ where almost degenerate in energy, and with remarkably similar Voronoi diagrams (see Fig.~2 of \cite{Calef15}).

\begin{figure}
\begin{center}
\includegraphics[width=12cm]{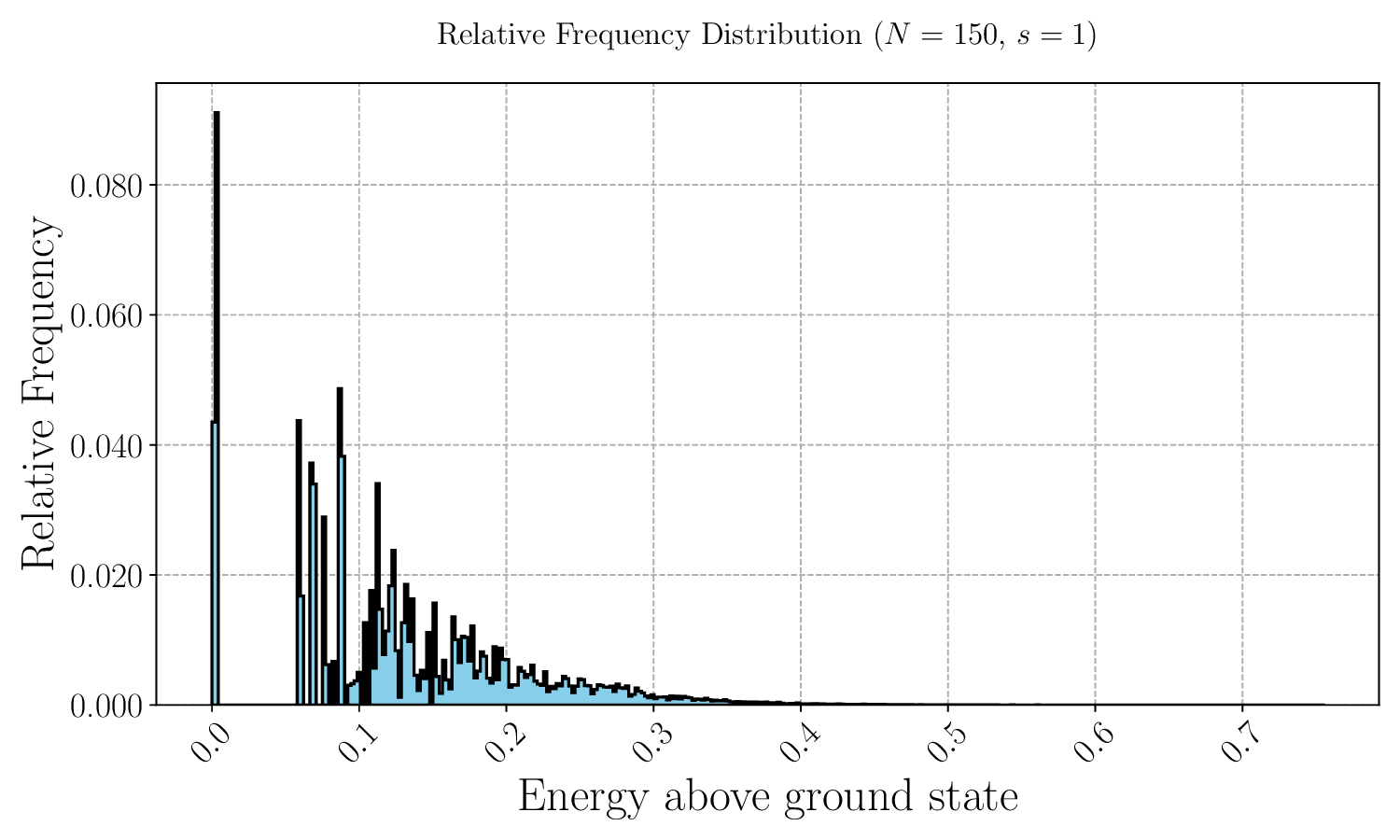}    
\caption{Histogram for the energies  for $N=150$ ($10^6$ random trials).}
\label{Fig_energy_frequency}
\end{center}
\end{figure}

\begin{figure}
\begin{center}
\includegraphics[width=12cm]{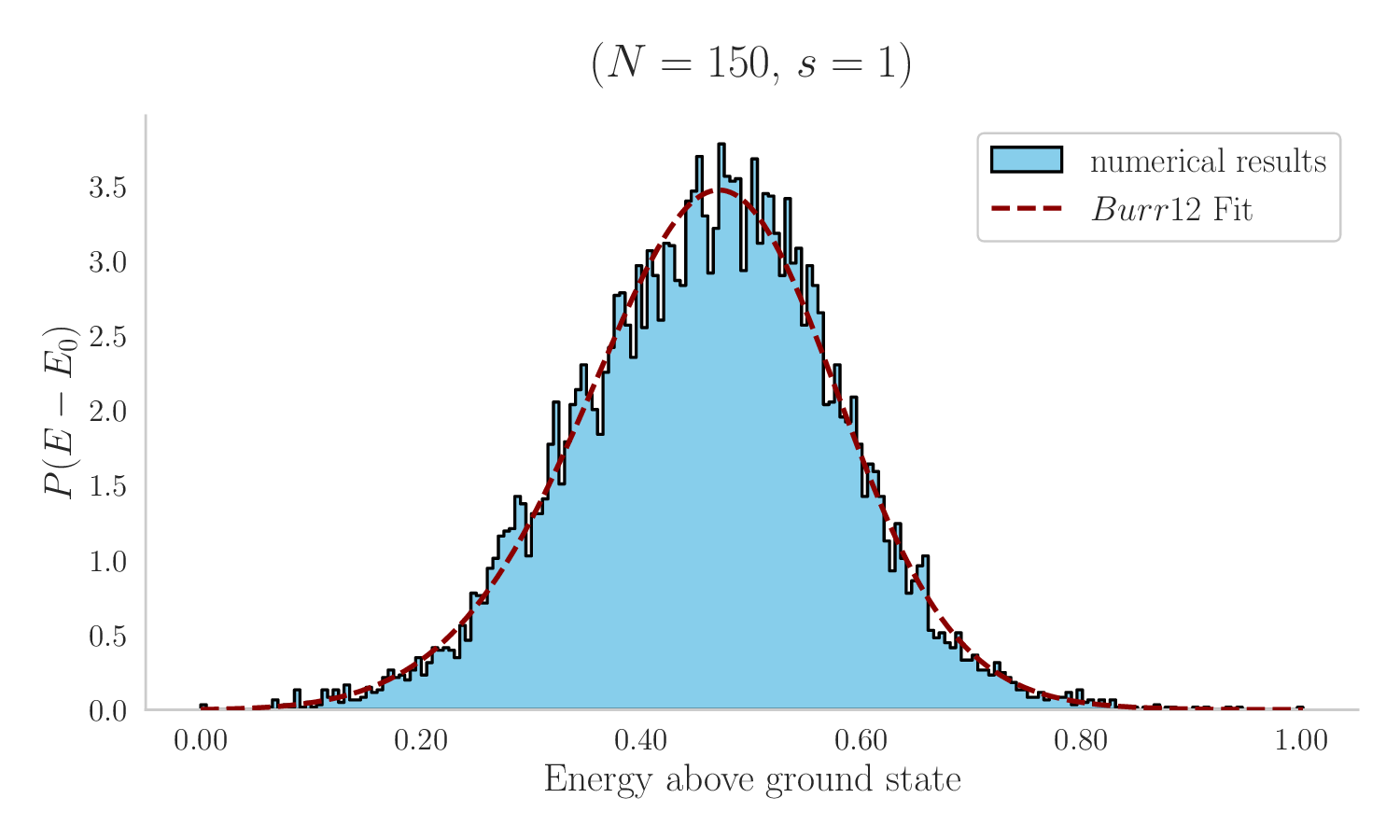}    
\caption{Histogram for the energies of independent configurations for $N=150$.}
\label{Fig_energy_distribution}
\end{center}
\end{figure}

In Fig.~\ref{Fig_energy_frequency} we plot the histogram of frequencies for configurations of $N=150$ charges obtained with $10^6$ {\sl random trials}~\footnote{The plot reports {\sl all} the trials without filtering repeated configurations.}:  this figure should be analyzed in conjunction with Fig.~\ref{Fig_energy_distribution} where we plot the probability distribution for configurations of $150$ charges, using only the independent configurations found in our experiments (and applying the different methods that we have described). 

Several notable observations emerge from these figures. First, the $10^6$ configurations shown in Fig.~\ref{Fig_energy_frequency} represent only about $50\%$ of the total number of distinct configurations we identified for $N = 150$. While the two lowest configurations appear very often, certain energy regions remain unpopulated---even though Fig.~\ref{Fig_energy_distribution} suggests that configurations should exist there. Most configurations are concentrated in the energy range between $0.3$ and $0.6$ above the ground state, yet Fig.~\ref{Fig_energy_frequency} accounts for only a small fraction of them. This discrepancy may help explain why the results of Erber and Hockney for $100 \leq N \leq 112$ fall short compared to ours---for instance, for $N = 112$, we identified $153$ configurations, whereas only $60$ were reported in~\cite{Erber97}. 

\begin{figure}
\begin{center}
\includegraphics[width=12cm]{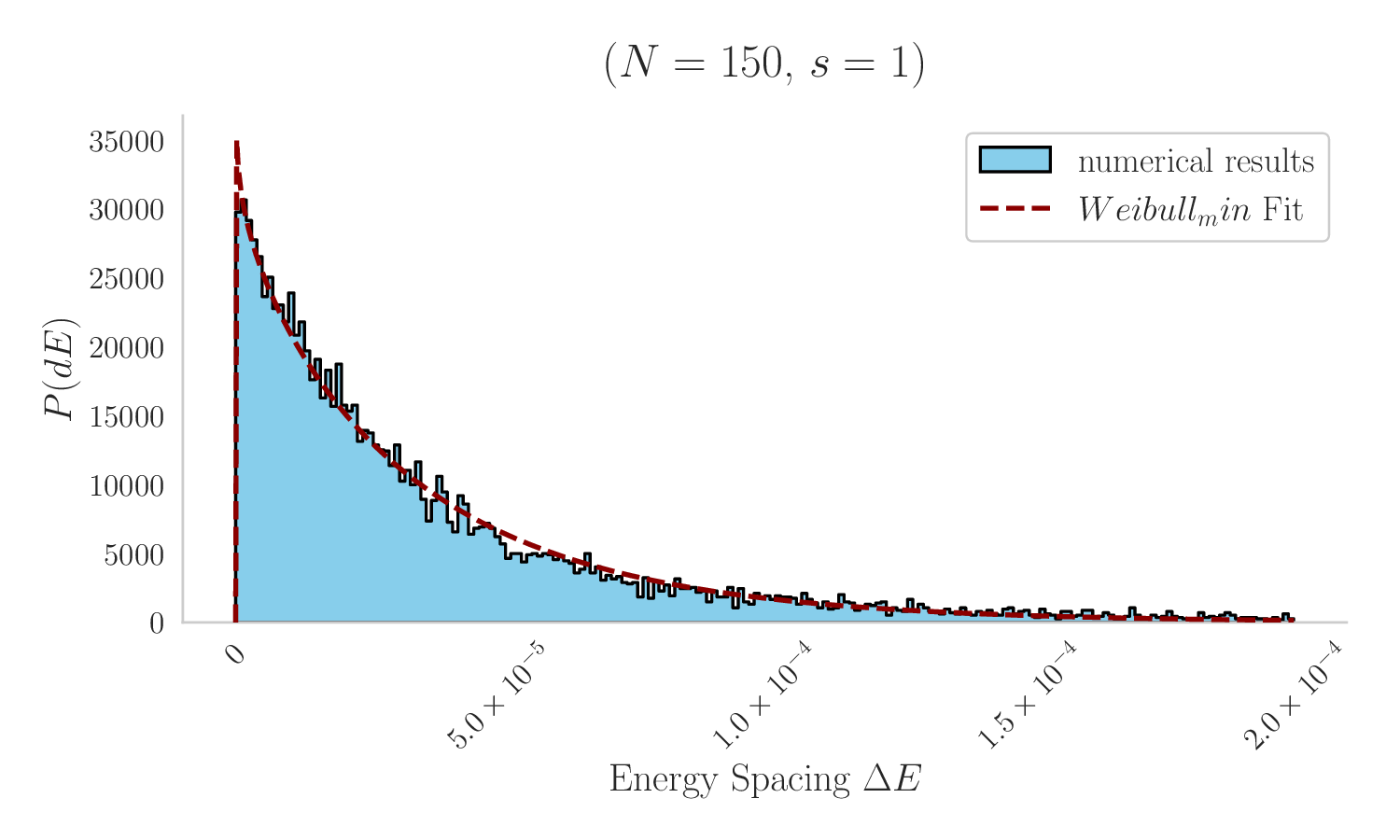}    
\caption{Histogram for the energy gaps (right)  for $N=150$.}
\label{Fig_denergy_distribution}
\end{center}
\end{figure}

Another interesting aspect is that, for the largest $N$ considered,  the energies of the independent configurations appear to be distributed following a Burr {\rm XII} distribution~\cite{Sanchez19} (see  Fig.~\ref{Fig_energy_distribution})
\begin{equation}
f(x;c,k,\lambda) = \frac{ck}{\lambda}\left(\frac{x}{\lambda}\right)^{c-1}\left[1 + \left(\frac{x}{\lambda}\right)^{c}\right]^{-(k+1)} \  ,
\end{equation}
where $c=6.36$, $k=4.39$ and $\lambda=0.777$. 

Similarly,  the energy gaps appear to follow a Weibull distribution  (Fig.~\ref{Fig_denergy_distribution}):
\begin{equation}
f(x;\lambda,k) = \frac{k}{\lambda}\left(\frac{x}{\lambda}\right)^{k-1}e^{-\left(\frac{x}{\lambda}\right)^k} \ , 
\end{equation}
where $k = 0.947$ and $\lambda = 3.530 \times 10^{-5}$. Weibull distribution is used  to describe  fragment size distribution~\cite{Tenchov86, Fang93, Lu02, Brouwers16, Fanfoni24}.

It is worth pointing out that Fig.~4 and 5 of ref.~\cite{Erber97} are the analogous of our Figs.~\ref{Fig_energy_distribution} and \ref{Fig_energy_frequency} for the joint case $N=111$ and $N=112$ (the two largest configurations studied by  Erber and Hockney).  Due to the limited number of configurations, however, 
the analysis carried out in ref.~\cite{Erber97} was not detailed.

\begin{figure}
\begin{center}
\includegraphics[width=10cm]{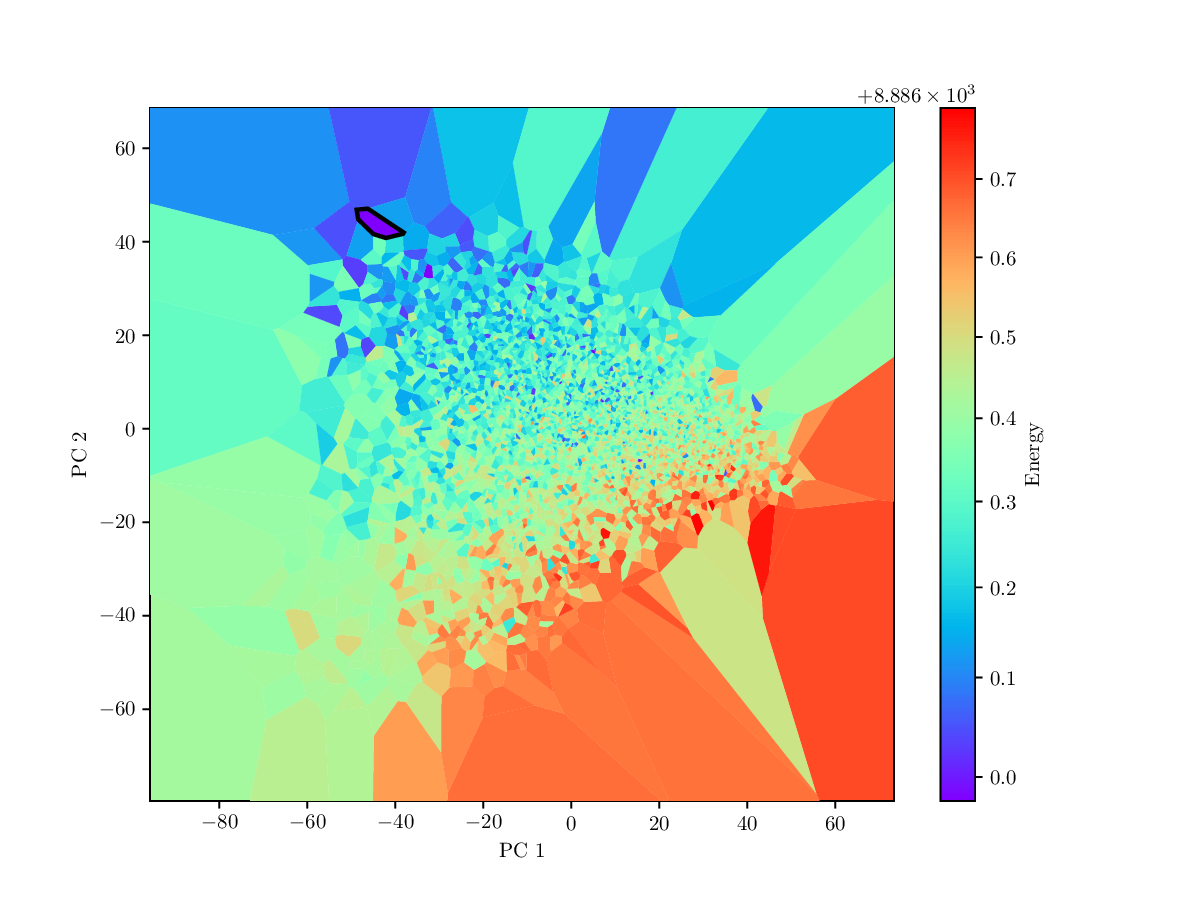}    
\caption{PCA for  local minima with 140 charges.}
\label{Fig_PCA_140}
\end{center}
\end{figure}

In Fig~\ref{Fig_PCA_140}  we have performed a partial component analysis (PCA) for the case of configurations with $N=140$:  the PCA uses the $\Delta$ distance between pairs of configurations to represent the configurations on a plane. $\Delta$ is the parameter that  corresponds to the maximum overlap between different configurations, that we have introduced in subsection \ref{sub_comp}. A similar analysis was carried out by one of us in ref.~\cite{Amore24} while studying the packing of congruent disks on a spherical cap.

Each configuration is represented as a point in the diagram and closeness between different points signals that the configurations are similar: while in the PCA of \cite{Amore24}
only the single points where plotted, here we represent the associated Voronoi diagrams, where the cells are colored according to the energy of the configuration the cell belongs to. In particular the border of the cell containing the global minimum is represented by a thick black line. 

This figure conveys important information: first, if the size of the Voronoi cells is large it  signals  that the corresponding configuration is relatively different from the remaining configurations (conversely, if the size is small, it is more likely that similar configurations could be found in proximity).  The question of how easily the ground state can be reached, which is relevant to  Smale's 7th problem (see \cite{Mehta15}) cannot be directly inferred from the diagram, because the probability of generating a given configuration starting 
from a random ansatz is not uniform: as a matter of fact, Erber and Hockney in Fig.5 of \cite{Erber97} modify their histogram of Fig.~4 showing the density of states, corrected to take into account the probability of occurrence of the single states. For the values of $N$ considered here (and in \cite{Erber97}) it is generally observed that identifying the global minimum is relatively easy. Even for larger configuration, one could find the global minimum either by performing a number of minimization trials on randomly generated ansatzes (that corresponds to the strategy of \cite{Erber97}) , or by generating an ansatz that once minimized corresponds to a point on the PCA diagram not far away from the target: by applying a number of perturbations to this configuration one can successfully land on the global minimum if the two are not far apart.

\subsection{Stationary states}
\label{sec:stat_states}

It appears that the solution landscape of the Thomson problem has not been exhaustively studied before: in fact, refs.~\cite{Mehta15} and \cite{Yin22} are
the only references that we have found where a (partial)  exploration has been carried out (Fig.2 of ref.~\cite{Yin22} regards the cases $N=5,7,9$). 

The exploration of the whole solution landscape is far more demanding than the exploration of the energy landscape, both because of the much larger number of 
configurations that are present and because finding these configurations requires appropriate approaches (such as those described in \cite{Mehta15, Yin22}).
The approach that we have adopted is different from those of \cite{Mehta15, Yin22}  because it converts the problem of finding the stationary points of the Thomson problem
to  a minimization problem of a suitable functional, whose (degenerate) global minima are the stationary configurations.

\begin{figure}
\begin{center}
\includegraphics[width=10cm]{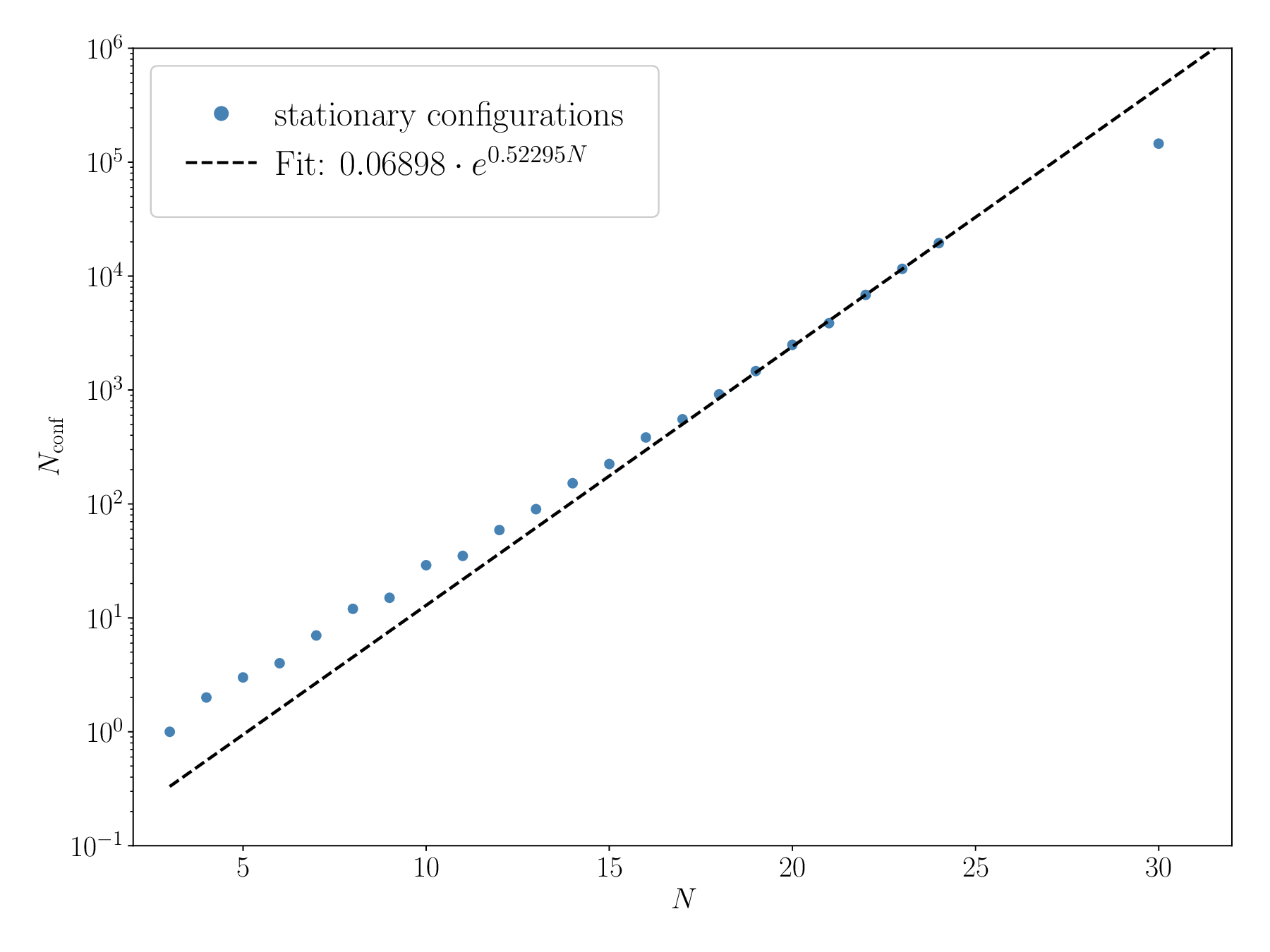}    
\caption{Number of stationary configurations for the Thomson problem.}
\label{Fig_N_stationary}
\end{center}
\end{figure}

We have conducted  a large number of numerical experiments for $2 \leq N \leq 24$, with the additional case $N=30$.
In Fig.~\ref{Fig_N_stationary} we report the number of stationary configurations found vs $N$ for the different cases studied (blue dots); the dashed line in the plot 
is the exponential fit over the range $15 \leq N \leq 24$\begin{equation}
N_{\rm conf} \approx 0.06898 \times e^{0.52295 \ N} \ .
\end{equation}

The exploration for $N=30$ has been partial and further numerical experiments would still be needed: despite having identified $145404$ configurations, the fit tells us that 
the true number of configurations is much larger ($N_{\rm conf}(30) > 4.4 \times 10^4$, using the fit).

\begin{figure}
\begin{center}
\includegraphics[width=10cm]{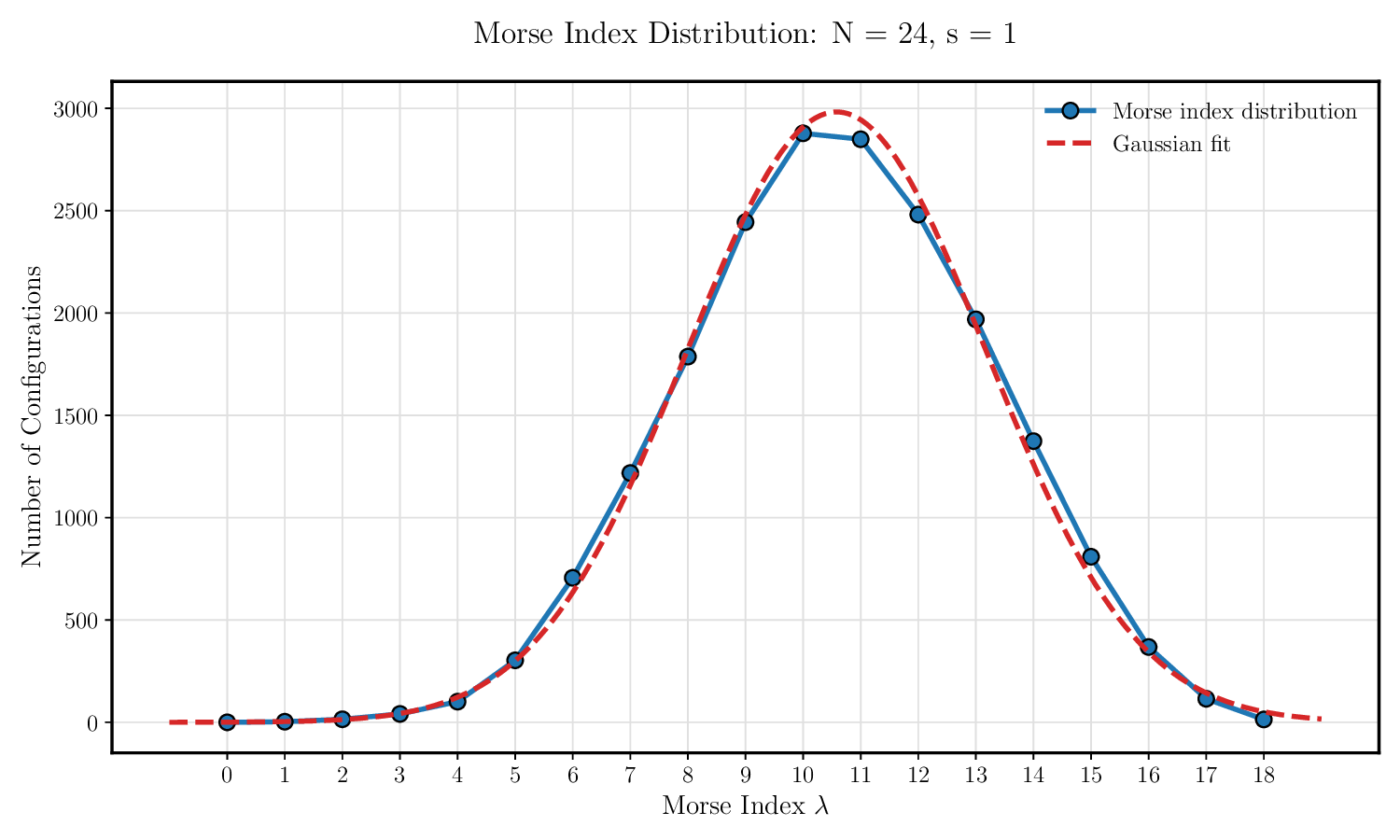}    
\caption{Number of stationary configurations for the Thomson problem.}
\label{Fig_histogram_index}
\end{center}
\end{figure}

In Fig.~\ref{Fig_histogram_index}  we plot the number of configurations found for $N=24$ as function of the Morse index (number of negative eigenvalues of the hessian) 
and observe  that this behavior is nicely described by a gaussian fit (similarly for the other cases that we have studied, with the peak of the gaussian slightly below $N/2$) in agreement with general considerations put forward by  Wales and Doye~\cite{Wales03}.

\begin{figure}
\begin{center}
\includegraphics[width=10cm]{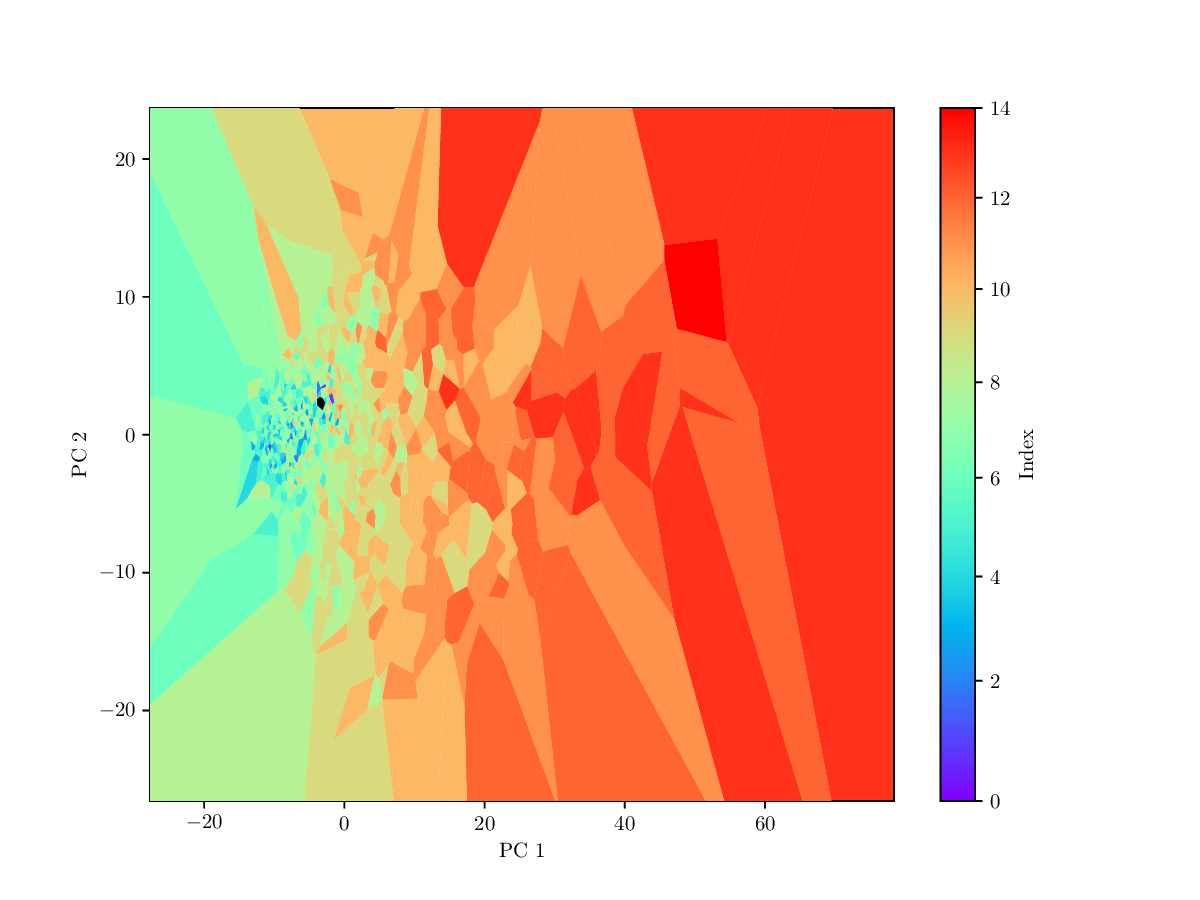}    
\caption{PCA for stationary states with 18 charges.}
\label{Fig_PCA_18}
\end{center}
\end{figure}

In analogy to what was done for local minima, we also used principal component analysis (PCA) to generate a two--dimensional 
representation of the solution landscape:  Fig.~\ref{Fig_PCA_18} illustrates the case for $N=18$,  where the cell with a thicker black border
denotes to the minimum; the remaining cells are colored according to the corresponding index. 
Notably, cells for high-index saddle points are typically larger than those for low-index saddle points, suggesting their configurations are more distinct.

\section{Conclusions}
\label{sec:concl}

In the present paper we have carried out an in--depth exploration of the energy and of the solution landscapes of the Thomson problem over a wide range of  $N$ (number of charges). 

This exploration allows to reach the following conclusions: 
\begin{itemize}
\item the number of local minima for several values of $N$ is substantially larger than what previously found by Erber and Hockney~\cite{Erber91, Erber95,Erber97} and by Calef et al.~\cite{Calef15}, suggesting that the exponential growth is much stronger than what previously estimated (to a lesser extent our results also improve the results of \cite{Mehta15}, that were conducted for selected values of $N$);
\item  the energy gaps  (both average and smallest) in a population of configurations of independent local minima  also decay exponentially with $N$, whereas the energy span grows linearly: the main consequence of this fact is that for sufficiently large $N$ the energy gap between consecutive configurations may be smaller than the precision achievable working in double precision;
\item configurations that are very similar in energy are not necessarily very similar themselves;
\item the distribution of energies and energy gaps appear to follow  a Burr 12 and a Weibull distributions respectively;
\item the number of stationary configurations grows exponentially with $N$, much more strongly than the number of local minima;
\item for a given $N$, the population of stationary points follows a gaussian distribution in the index, with mean slightly below $N/2$;
\end{itemize}

In addition to this, the demanding nature of this problem has forced us to devise more efficient algorithms for the exploration of the energy landscape: in particular
the upgrade and downgrade algorithms have proved to be very efficient both because they allow to generate a large spectrum of ansatzes and because such ansatzes typically correspond to more balanced initial configurations (therefore requiring less iterations to converge). 

On the software side, we have found the tools provided by python, scipy, numba, mpmath to be incredibly useful and powerful: as a matter of fact
 the largest configurations that we have studied in this paper  have required  several millions trials, a task that would have been out of range otherwise.

We feel that our work can be extended in multiple directions, the most immediate being:
\begin{itemize}
\item studying more general forms of potentials, particularly of the form of eq.~(\ref{eq_V}), including the logarithmic potential, relevant for Smale's seventh problem;
\item  extending this analysis to larger $N$, compatibly with the exponential growth of the number of configurations and with the extra need for precision
introduced by the progressive shrinking of the energy gaps;
\end{itemize}

\section*{Acknowledgements}
 The authors thank Dr. Jorge Muñoz for generously allows us to use his nodes in the University of Texas at El Paso's Jakar High Performance Computing system to support this research.
 The  research of P.A.  is supported by Sistema Nacional de Investigadores (M\'exico).

\end{document}